\documentclass[conference, compsoc]{IEEEtran}
\IEEEoverridecommandlockouts
\usepackage{subfigure}
\usepackage{caption}
\usepackage{hyperref}

\usepackage{amsmath}
\usepackage{booktabs}
\usepackage{graphicx}
\usepackage{multirow}
\usepackage{courier}
\usepackage{multirow}
\usepackage{acronym}
\usepackage{color}

\begin{document}

\title{Agent-based Simulation of Blockchains\footnotemark}

\author{
\IEEEauthorblockN{Edoardo Rosa}
\IEEEauthorblockA{Intuity, Padova, Italy\\
edoardo.rosa@studio.unibo.it}
\and
\IEEEauthorblockN{Gabriele D'Angelo, Stefano Ferretti }
\IEEEauthorblockA{Department of Computer Science and Engineering (DISI)\\University of Bologna, Italy\\
\{g.dangelo, s.ferretti\}@unibo.it}
}

\maketitle

\footnotetext{
The publisher version of this paper is available at \url{https://doi.org/10.1007/978-981-15-1078-6_10}.
\textbf{{\color{red} This is the pre-peer reviewed version of the article: ``Edoardo Rosa, Gabriele D'Angelo, Stefano Ferretti. Agent-based Simulation of Blockchains. Proceedings of the 19-th Asia Simulation Conference (AsiaSim 2019).''.}}}

\begin{abstract}
In this paper, we describe LUNES-Blockchain, an agent-based simulator of blockchains that is able to exploit Parallel and Distributed Simulation (PADS) techniques to offer a high level of scalability. To assess the preliminary implementation of our simulator, we provide a simplified modelling of the Bitcoin protocol and we study the effect of a security attack on the consensus protocol in which a set of malicious nodes implements a filtering denial of service (i.e.~Sybil Attack). The results confirm the viability of the agent-based modelling of blockchains implemented by means of PADS.\\
\end{abstract}


\section{Introduction}

Blockchain technologies are getting more and more hype these days, due to the vast range of possibilities of application in many distributed systems and networks~\cite{D'Angelo:2018,cryblock2019}. Traceability, auditing, attestation-as-a-service, regulation, cooperation, are just few examples of scenarios, other than the traditional fintech applications that made this technology famous.

\begin{figure}
  \centering
  \includegraphics[width=.8\linewidth]{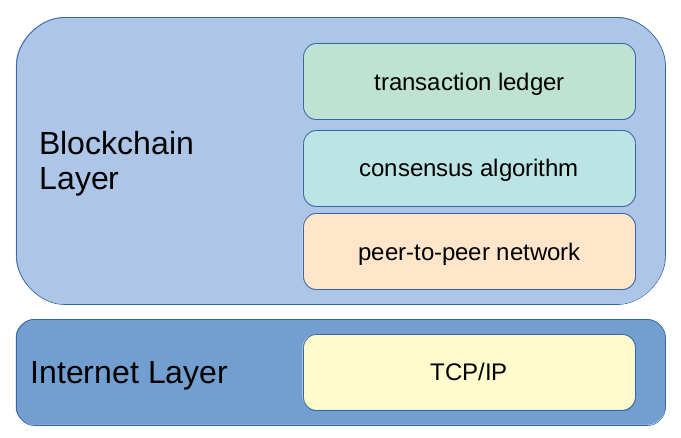}
  \caption{Blockchain protocol stack.}
  \label{fig:bc_stack}
\end{figure}

The blockchain can be treated as a protocol stack, in which each layer refers to a specific aspect of the blockchain. Figure \ref{fig:bc_stack} shows a simplified view of such blockchain protocol stack. At a coarse grained level of description, there are at least three main layers, on top of the Internet layer. The blockchain has an underlying peer-to-peer protocol, in charge of disseminating information on novel blocks being produced, to be added to the blockchain, or novel transactions that might be inserted into novel blocks. A flooding mechanism is often used to disseminate information, while the peer-to-peer overlay is built using some peer discovery mechanism~\cite{DBLP:journals/corr/abs-1801-03998}. For instance, a random selection protocol is used in Bitcoin, while Ethereum employs a UDP-based node discovery mechanism inspired by Kademlia~\cite{10.1007/3-540-45748-8_5}.
A consensus algorithm is used in order to let all nodes agree on the blockchain evolution. While the famous Bitcoin blockchain exploits a Proof-of-Work consensus scheme, many other possibilities exist, ranging from Proof-of-Stake and its plethora of variants, Proof-of-Authority, up to the Practical Byzantine Fault Tolerance consensus~\cite{DBLP:journals/corr/abs-1904-04098}. On top of the consensus layer, we have the transaction ledger, that records transactions and data. In the so called blockchain 2.0, i.e.~since Ethereum, these technologies offer the possibility to develop smart contracts, executed on the blockchain. A smart contract is a program representing an agreement that is automatically executable and enforceable by nodes that participate in the blockchain management. The execution of the program is triggered by transactions generated by an external account (i.e.~a user), and the program deterministically executes the terms of a contract, specified as software code~\cite{D'Angelo:2018}.

It is clear that, while this layered organization of a blockchain allows to isolate the very different aspects of this technology and allows obtaining a better understanding of the components' protocols, the functioning of each layer influences the performance of other layers. Thus, it becomes interesting to evaluate all possible alternatives of each component, and how possible modifications affect other aspects of the blockchain. However, the complexity of this technology, and the large scale nature of this distributed system make extremely difficult the evaluation process. In this sense, the simulation of the blockchain becomes an interesting evaluation strategy.

In this work, we present a novel blockchain simulator called LUNES-Blockchain that is able to exploit Parallel And Distributed Simulation (PADS) functionalities. The simulator mimics several functionalities of a blockchain, such as the peer-to-peer overlay management, its message dissemination scheme, the mining process (i.e.~the generation of a novel block) based on the generation of transactions. In particular, we show a specific implementation of the Bitcoin blockchain. Using the simulator, we then study a simple Denial-of-Service (DoS) attack (i.e.~Sybil Attack~\cite{Eyal:2018:MEB:3234519.3212998}) to the Bitcoin blockchain. As concerns the evaluation, in order to show the feasibility of the simulator, we provide some results related to different configurations of the DoS attack when applied to a simulated network composed of a large number of nodes.

The remainder of this paper is organized as follows. Section~\ref{sec:back} describes the background related to the paper subject. Section~\ref{sec:simulator} presents the blockchain simulator. The analysis of a DoS attack on the Bitcoin network is described in Section~\ref{sec:perf}. Finally, Section~\ref{sec:conc} provides some concluding remarks.

\section{Background and Related Work}\label{sec:back}

Blockchain is a technology that was initially proposed in the Bitcoin system, in 2009, by an anonymous author with the pseudo-name of Satoshi Nakamoto~\cite{Nakamoto_bitcoin}. In its first wave, the blockchain allowed parties to transact directly, i.e.~without any intermediary, by exchanging crypto money (Bitcoin) with confidence that no double spend was occurring. This was basically achieved using three key technologies: i) a globally shared ledger managed in a peer-to-peer fashion, ii) a mechanism for reaching consensus on the state of the ledger, and iii) immutability of the ledger and transactions (see Figure~\ref{fig:bc_stack}). All parties exploit pseudonyms, thus while transactions are known in the ledger, it is extremely difficult (even if not impossible, in some cases) to identify the involved parties, i.e.~the blockchain is pseudo-anonymous.

While Bitcoin basically allows for money transfers, a second wave of blockchain platforms followed, such as Ethereum, that enable other types of more complex applications. These platforms are based on the use of ``smart contracts'', that promote the development of decentralized applications, based on a Turing complete scripting language~\cite{Ethereum}. The execution of the public code composing the smart contract is carried out by the multiple nodes that are part of the system.

\subsection{The Peer-to-Peer Overlay and the Distributed Ledger}
Blockchains are distributed ledgers. To realize this, the system is organized as a peer-to-peer system, in which each participant has a copy of the shared ledger. This ledger records all the transactions within the blockchain. In Bitcoin, the ledger is a set of records of transactions that have occurred. Transactions are grouped in blocks. Each block contains a hash pointer to a previous block. It is this list of concatenated blocks that creates the blockchain.

Each novel transaction, generated by a node, is disseminated through the peer-to-peer system using a flooding dissemination protocol~\cite{gda-jpdc-2017,Ferretti20131631}. Such novel transaction will be considered, together with other not yet confirmed transactions, for the creation of a novel block. The generation of a block is based on a specific consensus scheme, which varies depending on the blockchain (see next section). Once a novel block has been generated, this block is disseminated in the overlay, through the same flooding dissemination scheme.

\subsection{Consensus Scheme}
The consensus scheme is in charge of ensuring that all nodes in the peer-to-peer system maintain the same view of the blockchain. Put in other words,  a mechanism is needed to allow all participants with copies of the ledger to come to consensus about the current state of the ledger and the uniqueness of transactions in the ledger. Several consensus schemes have been proposed in the past in the distributed system research area. Some of these schemes are utilized today in blockchain technologies (e.g.~Practical Byzantine Fault Tolerance~\cite{Castro:1999:PBF:296806.296824}). However, the principle approaches in blockchain are the Proof-of-Work (Pow, used in Bitcoin and the actual Ethereum) and Proof-of-Stake (PoS, used in the novel Ethereum version)~\cite{DBLP:journals/corr/abs-1904-04098}.

PoW works as follows.  Participants submit their transactions to the network. Nodes that participate to the transactions validation are called ``miners''. Miners verify that the submitted transactions are valid. Miners group these transactions into ``blocks''. Using this block as an input, the miners solve a computational crypto-puzzle that requires a large amount of computational power. When they solve the puzzle, they propagate the answer to other nodes along with the block of transactions. The other miners will accept the solution along with the block of transactions and add those transactions to the blockchain. 
As mentioned, a hash of this previous block in the chain is inserted in the novel block. This way, everyone can verify the ledger state in a tamper-proof manner.

The miners work is not for free.  A financial reward is assigned to the first node that solves the crypto-puzzle. PoW is often criticized for being a highly inefficient means of transacting, since the crypto-puzzle requires a tremendous waste of computation, hence causing a vast energy waste.

\subsection{Simulation of the Blockchain}
At the time of writing, literature on blockchain simulators is scarce. Usually, the main focus was on the analysis of the blockchain, the use of smart contracts and security issues. The typical approach is to develop smart contracts and test them using local blockchains. Remix, Metamask, Ganache, Multichain and the Ethereum test networks (e.g.~Ropsten, Rinkeby) are examples of environments thought to write, compile and debug smart contracts. In accordance to the multi-layered vision of a blockchain we discussed on the previous section, a common approach is to simulate just few aspects of a blockchain at a time.

In \cite{simblock}, a blockchain network simulator is presented. It is an event-driven simulator, that simulates the neighbor nodes selection of the peer-to-peer overlay. The mining activity is not simulated in detail, but a block generation is mimicked based on the computational capabilities of nodes.

In \cite{Gervais:2016}, the mining strategy of Bitcoin is simulated and studied. A network is modeled, but the propagation of transactions is not simulated, since the focal point is to study the impact of the block size, block interval, and the block request management system.

VIBES is a blockchain simulator, thought for large modeling scale peer-to-peer networks~\cite{Stoykov:2017}. The rationale behind this simulator is to provide a blockchain simulator that is not confined to the Bitcoin protocol, trying to provide support for large-scale simulations with thousands of nodes.

BlockSim is proposed as a Python framework to build discrete-event dynamic system models for blockchain systems~\cite{Alharby:2019}. BlockSim is organized in three layers: incentive layer, connector layer and system layer. Particular emphasis is given on the modeling and simulation of block creation through PoW.

In \cite{191667} is described a new methodology that enables the direct execution of multi-threaded applications inside of Shadow that is an existing parallel discrete-event network simulation framework. This is used to implement a new Shadow plug-in that directly executes the Bitcoin reference client software (i.e.~Shadow-Bitcoin).

At best of our knowledge, LUNES-Blockchain is the first simulator of blockchains that is able to take advantage of the performance speedup and extended scalability provided by PADS.

\section{Simulation of the Bitcoin Network}\label{sec:simulator}

With the aim to make this paper as much self-contained as possible, in this section we introduce some background on Discrete Event Simulation (DES) and Parallel And Distributed Simulation (PADS) techniques. After that, we describe the ART\`IS/GAIA simulation middleware and the LUNES simulation model that have been used for implementing LUNES-Blockchain.

\begin{table*}
\centering
\caption{Simulation and model parameters.}\label{table-perf}
\begin{tabular}{|l|c|l|}
\hline
Name & Value & Description \\
\hline
TTL                     & 16                    & Time-To-Live \\
DISSEMINATION           & 7                     & Dissemination protocol \\
PROBABILITY\_FUNCTION   & 2                     & Dissem. probability function \\
FUNC\_COEFF\_HIGHER     & 4                     & Dissem. high-order function coef. \\
FUNC\_COEFF\_LOWER      & 74                    & Dissem. low-order function coef. \\
END\_CLOCK              & 5000                  & Time-steps in each run \\
NODES                   & 10000                 & Number of nodes \\
MINERS\_COUNT           & 70\%                  & Percentage of miners \\
DIFFICULTY              & 6489747252517         & Difficulty value \\
HASHRATE                & 43983561622000000000  & Total hashrate (Hashes per sec.) \\
Network Topology        & random graph          & Topology of the Bitcoin network \\
Edges per Node          & 8                     & Number of edges per node \\
\hline
\end{tabular}
\end{table*}

DES is a simulation paradigm much appreciated for its usability and ability to model complex systems~\cite{FUJ00}. In a DES, a simulation model is represented through a set of state variables and the model evolution is modelled by the processing of events in chronological order. To respect the causality constraint of events in the real-world, each simulated event is timestamped (i.e.~occurs at a specific instant in the simulated time) and it represents a change of the state variables. Under the implementation viewpoint, the changes in the simulated system can be seen as the processing of an ordered sequence of timestamped events in the simulated model.

In a monolithic (i.e.~sequential) simulation, all the model state variables representing the simulated model are allocated in a single Physical Execution Unit (PEU) that is in charge of generating new events, managing the pending event list and processing the events that are extracted from the ordered list in timestamp order. This kind of simulator is very simple and it can be implemented using a single executing process. On the other hand, the simplicity of this simulator is often paid in terms of performance and scalability. For example, the scalability of the simulator is limited, both in terms of time required to complete the simulation runs and complexity of the system that can be modelled~\cite{1668384}.

An alternative approach, that is called PADS, is based on the parallelization/distribution of the simulator load. More specifically, a set of networked PEUs (e.g.~CPU cores, processors or hosts~\cite{gda-jpdc-2017,FUJ00} is in charge of executing the simulator. In this case, the simulation model is partitioned in a set of Logical Processes (LPs) that are executed on top of the PEUs that participate in the parallel/distributed simulation. Under the implementation viewpoint, each LP manages a local pending event list and the the events that need to be delivered to parts of the simulation model that are allocated in other PEUs are encapsulated in messages. The main advantage of PADS is that it enables the modelling and the processing of larger and more complex simulation models with respect to DES. On the other hand, the partitioning of the simulated model is not easy~\cite{gda-simpat-2017} and a synchronization algorithm among LPs is needed to guarantee the correct simulation execution~\cite{FUJ00}.

\subsection{ART\`IS/GAIA}
The \textit{Advanced RTI System} (ART\`IS) is a parallel and distributed simulation middleware in  which the simulation model is partitioned in a set of LPs. As previously described the parallel/distributed simulator is composed of interconnected PEUs and each PEU runs one or more LPs. The main service provided by ART\`IS to LPs is time management (i.e.~synchronization) that is necessary for obtaining correct simulation results in a parallel/distributed setup.

In a PADS, a relevant amount of execution time is spent in delivering the interactions between the model components. The means that the wall-clock execution time of PADS is highly dependent on the performance of the communication network (i.e.~latency, bandwidth and jitter) that connects the PEUs. It is obvious that reducing the communication overhead can speed up the simulator runs.

The \textit{Generic Adaptive Interaction Architecture} (GAIA) is a software layer built on top of ART\`IS~\cite{pads}. In GAIA, the simulation model is partitioned in a set of Simulated Entities (SEs) that can be seen as small model components. Each LP allocates some SEs and provides to them the basic simulation services (e.g.~synchronization and message passing). In other words, the simulated model behavior is obtained through the interactions among the SEs. Under the implementation viewpoint, the interactions are encapsulated by timestamped messages exchanged between the LPs. From the simulation modelling viewpoint, GAIA follows a Multi Agent System (MAS) approach in which each SE represents an agent.  In fact, each SE is an autonomous agent that is able to perform some specific actions (i.e.~implementing an individual behavior) and to interact with other agents in the simulation (i.e.~implementing group behaviors).

GAIA is able to reduce the communication overhead, that is common in PADS, clustering in the same LP the SEs that frequently interact together. In terms of communication overhead, clustering the heavily-interacting entities permits to reduce the amount of costly LAN/WAN/Internet communications that are replaced by efficient shared memory messages. In the current version of GAIA, the clustering of entities is based on a set of high-level heuristics that analyze the communication behavior in the simulation model without any knowledge of the specific simulation domain.

\subsection{LUNES}
LUNES (Large Unstructured NEtwork Simulator) is a simulator of complex networks implemented on top of ART\`IS/GAIA. The main goal of LUNES is to provide an easy-to-use tool for the modeling and simulation of interaction protocol on top of large scale unstructured graphs with different network topologies~\cite{gda-jpdc-2017}. The tool is implemented following a modular approach (i.e.~network creation, dissemination protocols definition, analysis of results) that facilitates its reuse. A main point of LUNES is that it is designed and implemented for PADS using the services provided by ART\`IS (i.e.~parallel and distributed processing) and GAIA (i.e.~adaptive self-clustering, dynamic computational and communication load-balancing). This permits the efficient simulation of very large scale models even in presence of a high-level of details in the modelled systems. The communication between nodes in the unstructured graphs is modelled in LUNES using a set of dissemination protocols that are based on gossip. The usage of LUNES is quite simple since it provides to the simulation modeller a high-level Application Programming Interface (API) for the implementation of the protocols to be simulated.

\subsection{LUNES-Blockchain}
LUNES-Blockchain is a simulation model based on LUNES that implements an agent-based representation of a generic blockchain. In LUNES-Blockchain, each node is represented by means of an agent that implements a local behavior and interacts with other agents. The representation of network nodes by means of agents simplifies the development of the model and, in our view, it adds a high-level of extensibility to the simulation model.

In this preliminary work, we assess if it possible to study, via simulation, some security issues common to blockchains. In particular, we are interested in the behavior of the Bitcoin blockchain. This aspect has influenced some of the design choices that have been taken and that will be discussed in the following of this section. The design of LUNES-Blockchain has been organized in steps: i) modeling and simulation of a generic blockchain; ii) modelling of the specific aspects of the Bitcoin blockchain; iii) modelling of a malicious filtering DoS attack on Bitcoin.

Since the implementation in LUNES of the dissemination mechanism used in Bitcoin (called Dandelion/Dandelion++) is currently under development, LUNES-Blockchain models the dissemination in the Bitcoin network using one of the gossip protocols already provided by LUNES (i.e.~the degree dependent dissemination algorithm \cite{gda-jpdc-2017}). The main effect of this choice is the better communication delay (i.e.~latency) provided by the degree dependent dissemination with respect to Dandelion (due to the absence of the anonymity phase implemented in Dandelion).

The next version of LUNES-Blockchain will support Dandelion and a more accurate representation of the Bitcoin network topology~\cite{7816866}. These modifications will be useful to study some specific behaviors of the Bitcoin network. For example, when studying scalability aspects that are related to the Bitcoin implementation.

A relevant aspect in the modelling of the Bitcoin blockchain is the network size. Given its dynamic nature, it not possible to identify a specific number of nodes, but estimates assert that the Bitcoin network size is about $10.400$ nodes~\cite{bitnodes}. It is worth noticing that not all active nodes in the network are miners. In fact, it is not mandatory that all nodes participate in the creation of new block to be added to the blockchain. On the other side, all active nodes participate in the reception, validation and broadcast of new blocks, thus maintaining updated their local copy of the blockchain. Due to how PoW has been designed, in Bitcoin the majority of miners are part of mining pools. In this work, we are not interested in modelling each miner that is part of a mining pool. This is due to the fact that the malicious behavior we are interested in, considers each mining pool as a single node. The current implementation of LUNES-Blockchain permits to define the percentage of simulated nodes acting as miners, and for each miner its specific hashrate. The hashrate is defined as the speed at which a processing unit is able to complete the hash operations that are used to solve the cryptopuzzle of the PoW.
In LUNES-Blockchain, the simulation of the mining process is modelled to respect the difficulty and behavior of the Bitcoin mining process, but without the computational overhead caused by the real implementation of PoW. 

Another relevant issue is the modeling of time in the simulation. The current behavior of the Bitcoin network is to create and publish a new block every 10 minutes. On the other hand, the simulation implements a time-stepped synchronization algorithm~\cite{FUJ00} in which the simulated time is dived in a sequence of time-steps. In the current implementation of LUNES-Blockchain, each time-step represents one minute of simulated time. This means that, on the average, every 10 time-steps a new simulated block is created and propagated to the whole network by means of the gossip-based dissemination protocol.

LUNES-Blockchain is available for peer-review and it will be included in the forthcoming release of LUNES that will be available in source code format on the research group website~\cite{pads}.

\section{Evaluation of a DoS Attack}\label{sec:perf}

\begin{figure*}
  \centering
  \includegraphics[width=0.7\linewidth]{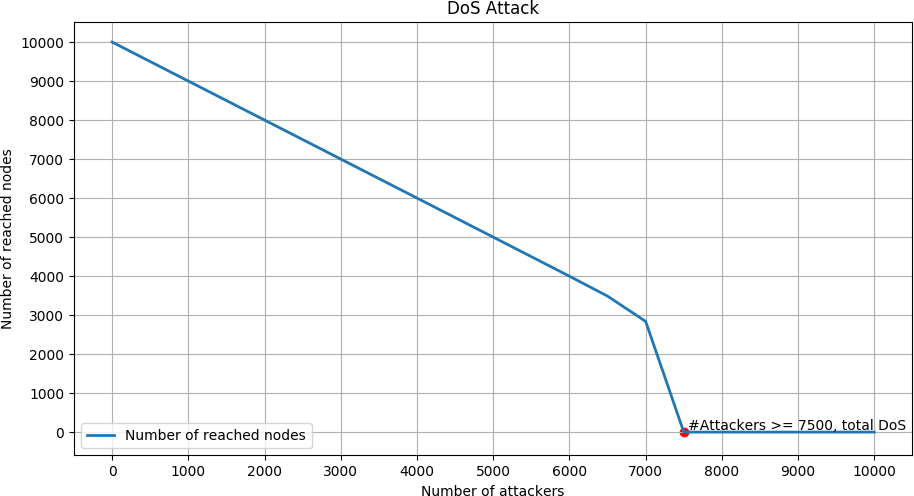}
  \caption{Average number of reached nodes during a DoS filtering attack.}
  \label{fig:attackDOS}
\end{figure*}

\begin{figure*}
  \centering
  \includegraphics[width=0.7\linewidth]{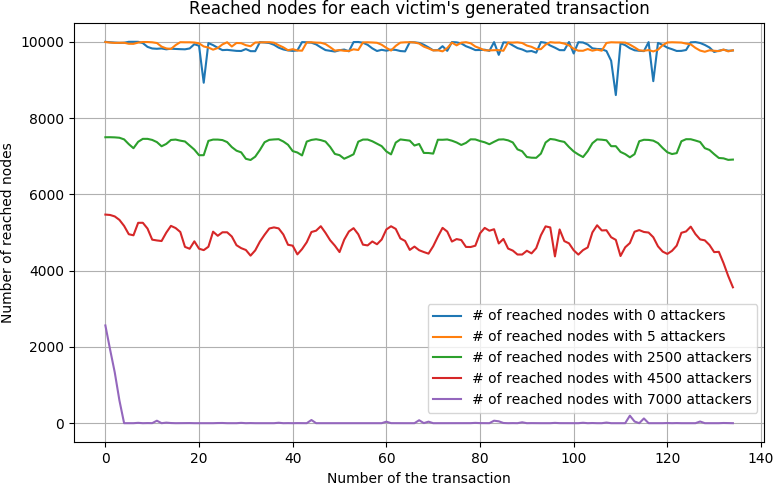}
  \caption{Number of reached nodes for each transaction during a DoS filtering attack. Different setups with an increasing number of attacking nodes.}
  \label{fig:attackDOSmix}
\end{figure*}

In this paper we investigate the modelling and simulation of a well-known type of DoS attack on the Bitcoin network.  To foster the reproducibility of our experiments, all the parameters used to setup up LUNES-Blockchain are described in Table~\ref{table-perf}. The parameters reported in the table mimic the Bitcoin network as it resulted in the fourth quarter of 2018.

The results shown in the following section have been obtained running the simulator on an Intel i7-6700 3.40GHz with 16GB of RAM and Arch Linux (x64) as operating system. The execution of a single simulation run of LUNES-Blockchain with the model parameters described in Table~\ref{table-perf} and using a single LP (i.e.~sequential simulation) requires an average of 60 seconds. This means that LUNES-Blockchain is quite efficient in the simulation of limited size blockchains even in a sequential (i.e.~monolithic) setup. More populated networks, the modelling of more complex attacks and the simulation of blockchains supporting Smart Contracts will benefit of the PADS approach provided by ART\`IS/GAIA. In future works, the preliminary evaluation reported in this paper, it will be followed by a full-fledged validation and scalability assessment of the simulator.

The attack that has been implemented is a filtering DoS in which a set of malicious nodes silently drop all messages that originated from a given node. This kind of filtering evolves in a Sybil Attack when the attacked node is completely surrounded by attackers. In other words, we simulate the condition in which the attacked node is totally unable to communicate with the rest of the network, with the effect that all its mining outcomes and transactions are discarded by the malicious sybils, and none among honest nodes in the blockchain overlay receive them. This specific experiment has been implemented as a sequence of simulation runs. We varied the number of malicious nodes from $1$ up to $9999$. With $9999$ attackers, the network is all made of malicious nodes, with the exception of the attacked one. For every run, the average number of nodes reached by each message originated from the attacked node has been calculated. 

It is worth noticing that this kind of evaluation, under the simulation viewpoint, is quite costly. In fact, many simulator runs have to be executed both for exploring all the different configurations and for obtaining statistically significant results. In other words, the efficiency of the simulator (i.e.~execution speed) is of main importance. The results that have been obtained are reported in Figures~\ref{fig:attackDOS} and \ref{fig:attackDOSmix}. The outcomes reported in the figures are comparable with theoretical results expected for this kind of attack against peer-to-peer botnets~\cite{botnets}. Clearly, this does not represent a validation of the proposed simulation model but it is a positive outcome.

Figure~\ref{fig:attackDOS} shows that the average number of nodes reached by the messages proportionally decreases with the increase of attackers. When the number of attackers is larger than $7000$, there is a sharp decrease in the number of nodes reached by the messages. With more than $7500$ attackers, the Sybil Attack is complete and the attacked node is disconnected from the network.

Figure~\ref{fig:attackDOSmix} shows the number of reached nodes when considering up to $140$ transactions that are one-by-one disseminated in the network. In the figure, it is possible to see the effect of an increasing number of malicious nodes on each transaction that is delivered in the network. When there are no attackers (or a few of them), almost every transaction obtains a complete broadcast. When we increase the number of malicious nodes, then the filtering effect is evident on the number of reached nodes.

\section{Conclusions}\label{sec:conc}

In this paper, we have introduced a new agent-based simulator called LUNES-Blockchain for the simulation of large scale and complex blockchains. LUNES-Blockchain has been used for implementing a preliminary model of the Bitcoin network and to study a simple filtering Denial-of-Service that is usually called Sybil Attack. To the best of our knowledge, this is the first blockchain simulator that is able to exploit the performance speedup and improved scalability offered by Parallel and Distributed Simulation (PADS).

As a future work, we plan to perform a more extended validation of LUNES-Blockchain, to improve the accuracy the Bitcoin model implementing the Dandelion/Dandelion++ dissemination protocol and to consider a more accurate topology of the Bitcoin network. On the other hand, LUNES-Blockchain can be extended to model blockchains that are capable of Smart Contracts execution (e.g.~Ethereum). Finally, we plan to investigate other common attacks on the network that are based on the presence of malicious nodes.

\small{
\bibliographystyle{abbrv}
\bibliography{biblio}  

\begin{thebibliography}{10}

\bibitem{Alharby:2019}
M.~Alharby and A.~van Moorsel.
\newblock Blocksim: A simulation framework for blockchain systems.
\newblock {\em SIGMETRICS Perform. Eval. Rev.}, 46(3):135--138, Jan. 2019.

\bibitem{simblock}
Y.~Aoki, K.~Otsuki, T.~Kaneko, R.~Banno, and K.~Shudo.
\newblock Simblock: {A} blockchain network simulator.
\newblock In {\em Proc. of the 2nd Workshop on Cryptocurrencies and Blockchains
  for Distributed Systems}, CryBlock'19. IEEE, 2019.

\bibitem{bitnodes}
Bitnodes.
\newblock {Global Bitcoin Nodes Distribution}.
\newblock \url{https://bitnodes.earn.com/}, 2019.

\bibitem{Ethereum}
V.~Buterin.
\newblock A next-generation smart contract and decentralized application
  platform.
\newblock White Paper, 2018.
\newblock \url{https://github.com/ethereum/wiki/wiki/White-Paper}, Accessed on
  2018-03-02.

\bibitem{Castro:1999:PBF:296806.296824}
M.~Castro and B.~Liskov.
\newblock Practical byzantine fault tolerance.
\newblock In {\em Proceedings of the Third Symposium on Operating Systems
  Design and Implementation}, OSDI '99, pages 173--186, Berkeley, CA, USA,
  1999. USENIX Association.

\bibitem{gda-jpdc-2017}
G.~D'Angelo and S.~Ferretti.
\newblock Highly intensive data dissemination in complex networks.
\newblock {\em Journal of Parallel and Distributed Computing}, 99:28 -- 50,
  2017.

\bibitem{pads}
G.~D'Angelo and S.~Ferretti.
\newblock {Parallel And Distributed Simulation (PADS) Research Group}.
\newblock \url{http://pads.cs.unibo.it}, 2019.

\bibitem{D'Angelo:2018}
G.~D'Angelo, S.~Ferretti, and M.~Marzolla.
\newblock A blockchain-based flight data recorder for cloud accountability.
\newblock In {\em Proc. of the 1st Workshop on Cryptocurrencies and Blockchains
  for Distributed Systems}, CryBlock'18, pages 93--98, New York, NY, USA, 2018.
  ACM.

\bibitem{gda-simpat-2017}
G.~D’Angelo.
\newblock The simulation model partitioning problem: an adaptive solution based
  on self-clustering.
\newblock {\em Simulation Modelling Practice and Theory (SIMPAT)}, 70:1 -- 20,
  2017.

\bibitem{1668384}
E.~Egea-Lopez, J.~Vales-Alonso, A.~Martinez-Sala, P.~Pavon-Mario, and
  J.~Garcia-Haro.
\newblock Simulation scalability issues in wireless sensor networks.
\newblock {\em Communications Magazine, IEEE}, 44(7):64 -- 73, july 2006.

\bibitem{Eyal:2018:MEB:3234519.3212998}
I.~Eyal and E.~G. Sirer.
\newblock Majority is not enough: Bitcoin mining is vulnerable.
\newblock {\em Commun. ACM}, 61(7):95--102, June 2018.

\bibitem{Ferretti20131631}
S.~Ferretti.
\newblock Gossiping for resource discovering: An analysis based on complex
  network theory.
\newblock {\em Future Generation Computer Systems}, 29(6):1631 -- 1644, 2013.

\bibitem{FUJ00}
R.~Fujimoto.
\newblock {\em Parallel and Distributed Simulation Systems}.
\newblock {\it Wiley \& Sons}, 2000.

\bibitem{DBLP:journals/corr/abs-1801-03998}
A.~E. Gencer, S.~Basu, I.~Eyal, R.~van Renesse, and E.~G. Sirer.
\newblock Decentralization in bitcoin and ethereum networks.
\newblock {\em CoRR}, abs/1801.03998, 2018.

\bibitem{Gervais:2016}
A.~Gervais, G.~O. Karame, K.~W\"{u}st, V.~Glykantzis, H.~Ritzdorf, and
  S.~Capkun.
\newblock On the security and performance of proof of work blockchains.
\newblock In {\em Proceedings of the 2016 ACM SIGSAC Conference on Computer and
  Communications Security}, CCS '16, pages 3--16, New York, NY, USA, 2016. ACM.

\bibitem{10.1007/3-540-45748-8_5}
P.~Maymounkov and D.~Mazi{\`e}res.
\newblock Kademlia: A peer-to-peer information system based on the xor metric.
\newblock In P.~Druschel, F.~Kaashoek, and A.~Rowstron, editors, {\em
  Peer-to-Peer Systems}, pages 53--65, Berlin, Heidelberg, 2002. Springer
  Berlin Heidelberg.

\bibitem{191667}
A.~Miller and R.~Jansen.
\newblock Shadow-bitcoin: Scalable simulation via direct execution of
  multi-threaded applications.
\newblock In {\em 8th Workshop on Cyber Security Experimentation and Test
  ({CSET} 15)}, Washington, D.C., Aug. 2015. {USENIX} Association.

\bibitem{Nakamoto_bitcoin}
S.~Nakamoto.
\newblock Bitcoin: A peer-to-peer electronic cash system,”
  http://bitcoin.org/bitcoin.pdf, 2009.

\bibitem{7816866}
T.~{Neudecker}, P.~{Andelfinger}, and H.~{Hartenstein}.
\newblock Timing analysis for inferring the topology of the bitcoin
  peer-to-peer network.
\newblock In {\em 2016 Intl IEEE Conferences on Ubiquitous Intelligence
  Computing, Advanced and Trusted Computing, Scalable Computing and
  Communications, Cloud and Big Data Computing, Internet of People, and Smart
  World Congress}, pages 358--367, July 2016.

\bibitem{Stoykov:2017}
L.~Stoykov, K.~Zhang, and H.-A. Jacobsen.
\newblock Vibes: Fast blockchain simulations for large-scale peer-to-peer
  networks: Demo.
\newblock In {\em Proceedings of the 18th ACM/IFIP/USENIX Middleware
  Conference: Posters and Demos}, Middleware '17, pages 19--20, New York, NY,
  USA, 2017. ACM.

\bibitem{botnets}
A.~L. Verigin.
\newblock {Evaluating the Effectiveness of Sybil Attacks Against Peer-to-Peer
  Botnets}.
\newblock \url{http://hdl.handle.net/1828/5095}, 2018.

\bibitem{DBLP:journals/corr/abs-1904-04098}
Y.~Xiao, N.~Zhang, W.~Lou, and Y.~T. Hou.
\newblock A survey of distributed consensus protocols for blockchain networks.
\newblock {\em CoRR}, abs/1904.04098, 2019.

\bibitem{cryblock2019}
M.~Zichichi, M.~Contu, S.~Ferretti, and G.~D'Angelo.
\newblock Likestarter: a smart-contract based social dao for crowdfunding.
\newblock In {\em Proc. of the 2nd Workshop on Cryptocurrencies and Blockchains
  for Distributed Systems}, CryBlock'19. IEEE, 2019.

\end{thebibliography}
}

\end{document}